\def\submission{0}
\def\fullversion{1}
\def\todos{0}

\ifnum\fullversion=0
\ifnum\submission=0
\documentclass[manuscript]{acmart}
\else
\documentclass[manuscript,review]{acmart}
\fi
\else
\documentclass[11pt]{llncs}
\fi

\ifnum\fullversion=0
\usepackage{libertine}
\fi
\ifnum\fullversion=1
\usepackage{fullpage}
\fi
\usepackage{todonotes}
\usepackage{url}
\usepackage{hyperref}
\usepackage{float}
\usepackage{subcaption}
\usepackage{wrapfig}
\usepackage{multirow}
\usepackage{graphicx}
\ifnum\todos=1
\newcommand{\tore}[1]{{\color{purple}Tore: #1}}
\else
\newcommand{\tore}[1]{}
\fi

\newcommand{\servers}{\ensuremath{n}}
\newcommand{\server}{\ensuremath{\mathcal{S}}}
\newcommand{\client}{\ensuremath{\mathcal{C}}}
\newcommand{\reader}{\ensuremath{\mathcal{R}}}

\newcommand{\pp}{\ensuremath{\textit{pp}}}
\newcommand{\aux}{\ensuremath{\textit{aux}}}
\newcommand{\pkissuer}{\ensuremath{\pk_{\server}}}
\newcommand{\pkclient}{\ensuremath{\pk_{\client}}}
\newcommand{\skclient}{\ensuremath{\sk_{\client}}}
\newcommand{\pkse}{\ensuremath{\pk_\SE}}
\newcommand{\skse}{\ensuremath{\sk_\SE}}

\newtheorem{construction}{Construction}

\newcommand{\Hash}{\mathcal{H}}

\newcommand{\pk}{\mathit{pk}}
\newcommand{\sk}{\mathit{sk}}
\newcommand{\bk}{\mathit{bk}}

\newcommand{\pwd}{\mathit{pw}}

\newcommand{\uid}{\mathit{uid}}
\newcommand{\vid}{\mathit{vid}}

\newcommand{\SE}{\texttt{SE}}

\newcommand{\reg}{\mathsf{reg}}

\newcommand{\DSIG}{\mathsf{DSIG}}

\newcommand{\SIG}{\mathsf{SIG}}

\newcommand{\kgen}{\mathsf{KGen}}
\newcommand{\blind}{\mathsf{Blind}}
\newcommand{\unblind}{\mathsf{Unblind}}

\newcommand{\comb}{\mathsf{Comb}}
\newcommand{\sign}{\mathsf{Sign}}

\newcommand{\refresh}{\mathsf{Refresh}}

\newcommand{\vf}{\mathsf{Vf}}

\renewcommand{\gets}{\ensuremath{\leftarrow}}
\renewcommand{\to}{\ensuremath{\rightarrow}}
\newcommand{\getsr}{\leftarrow_\textrm{\tiny R}}
\newcommand{\tor}{\rightarrow_\textrm{\tiny R}}

\newcommand{\zo}{\{0,1\}}

\newsavebox{\fboxenvbox}
\newenvironment{fboxenv}
    {\begin{lrbox}{\fboxenvbox}}
    {\end{lrbox}\fbox{\usebox{\fboxenvbox}}}

\def\veclen{20}                  

\newcommand{\OneRoundProt}[1][\veclen]{
  \unitlength1ex 
  \begin{picture}(16,0)
    \put(0,0){\line(1,0){12}} 
    \put(12,0){\line(0,-1){4}}
    \put(12,-2){\line(1,0){4}}
    \put(12,-4){\vector(-1,0){12}} 
  \end{picture}}

\ifnum\fullversion=0
\AtBeginDocument{%
  \providecommand\BibTeX{{%
    \normalfont B\kern-0.5em{\scshape i\kern-0.25em b}\kern-0.8em\TeX}}}
\copyrightyear{2021} 
\acmYear{2021} 
\setcopyright{acmlicensed}\acmConference[ARES 2021]{The 16th International Conference on Availability, Reliability and Security}{August 17--20, 2021}{Vienna, Austria}
\acmBooktitle{The 16th International Conference on Availability, Reliability and Security (ARES 2021), August 17--20, 2021, Vienna, Austria}
\acmPrice{15.00}
\acmDOI{10.1145/3465481.3469212}
\acmISBN{978-1-4503-9051-4/21/08}
\fi

\begin{document}

\title{A Holistic Approach to Enhanced Security and \\Privacy in Digital Health Passports}

\ifnum\submission=0
\ifnum\fullversion=0
\author{Tore Kasper Frederiksen}
\authornotemark[1]
\email{tore.frederiksen@alexandra.dk}
\affiliation{%
  \institution{The Alexandra Institute}
  \streetaddress{Aabogade 34}
  \city{Aarhus}
  \country{Denmark}
  \postcode{8200}
}
\else
\author{Tore Kasper Frederiksen\inst{1}}
\institute{Security Lab, Alexandra Institute, {\sc Denmark}\thanks{This is a preprint. Original version at doi: \url{https://doi.org/10.1145/3465481.3469212}}\\
\email{tore.frederiksen@alexandra.dk}}
\fi
\fi


\ifnum\submission=0
\ifnum\fullversion=0
\renewcommand{\shortauthors}{Frederiksen}
\fi
\fi



\ifnum\fullversion=1
\maketitle
\fi
\begin{abstract}
As governments around the world decide to deploy digital health passports as a tool to curb the spread of Covid-19, it becomes increasingly important to consider how these can be constructed with privacy-by-design.

In this paper we discuss the privacy and security issues of common approaches for constructing digital health passports.
We then show how to construct, and deploy, secure and private digital health passports, in a simple and efficient manner.
We do so by using a protocol for distributed password-based token issuance, secret sharing and by leveraging modern smart phones' secure hardware.

Our solution only requires a constant amount of asymmetric cryptographic operations and a single round of communication between the user and the party verifying the user's digital health passport, and only two rounds between the user and the server issuing the digital health passport.
\end{abstract}
\ifnum\fullversion=0
\maketitle
\fi
\keywords{Digital Heath Passports, Distributed Signatures, Distributed Cryptography, Protocol Deployment}

\section{Introduction}
\label{sec:intro}
In the beginning of 2020, the world as we knew it changed. Within the period of a few months, a novel disease, Covid-19, spread to Europe, the US, and soon the rest of the world. 
Covid-19 turned out to be infectious and deadly enough to require almost everyone in the world to change their daily habits and interactions in order to limit its continuous rampage. 
Technologists quickly offered their help with this crisis. 
New smart phone apps, known as \emph{contact tracing} apps, were developed to help in stopping the spread of the disease.
The purpose of these apps is to log the user's close proximity with other users or locations. 
If a location or a user had been associated with an outbreak, it would then be possible to notify everyone who had been in their close proximity.
As soon as the idea of contact tracing apps was even suggested, privacy advocates were quick to argue how they could be used for mass surveillance and tracking.
Fortunately, security experts and cryptographers~\cite{dp3t}, along with Apple and Google~\cite{appleGoogle}, quickly got to work to make privacy-by-design frameworks for contact tracing. 
These apps use cryptographic techniques to ensure that the company or government hosting them learn as little as possible about the users' movements and proximity to one another. 
They showed that it is possible to use advanced and pervasive technologies in disease prevention \emph{without} compromising the privacy or anonymity of the users~\cite{dp3t,contactSecurity}.
Furthermore, the whole debate about contact tracing also showed extensive academic concern about privacy~\cite{letter}, in the setting where misuse could lead to mass government tracking of basically all citizens.

\subsection{Digital Health Passports}
Despite limited efficacy and large infrastructure overhead, most agree that contact tracing is an important tool in the fight against disease spreading~\cite{Keeling2020,AP21}. 
Fortunately, this is not the only technological tool used to fight Covid-19. 
Medical companies and universities were quick to start the development of vaccines. 
These vaccines were developed, tested, produced and rolled out at an astonishing time frame of less than a year~\cite{Polack20,Baden21}!
However, rolling out these vaccines to the entire world will, unfortunately, most likely take years. 
For this reason, testing remains a powerful tool in the prevention of disease spreading. 
On top of this, due to the recent deployment of the vaccines, it is not yet clear how long immunity will last. Something that might also depend on \emph{which} vaccine has been given.
For this reason, many governments around the world are using – or planning on using – so-called \emph{Immunity} or \emph{Digital Health Passports} (DHP). 

The general idea of a DHP is that a time-constrained certification is issued to a user which certifies that the user is most likely not contagious. 
A DHP can be issued either based on vaccination status, test status or verified immunity~\cite{eu}. 
They can come in either the shape of a digitally certified physical document, but due to the highly time-constrained nature of its validity (in particular in relation to testing), an on-demand solution is more applicable.

\begin{wrapfigure}{r}{0.45\textwidth}
	\centering
	\includegraphics[width=0.4\textwidth]{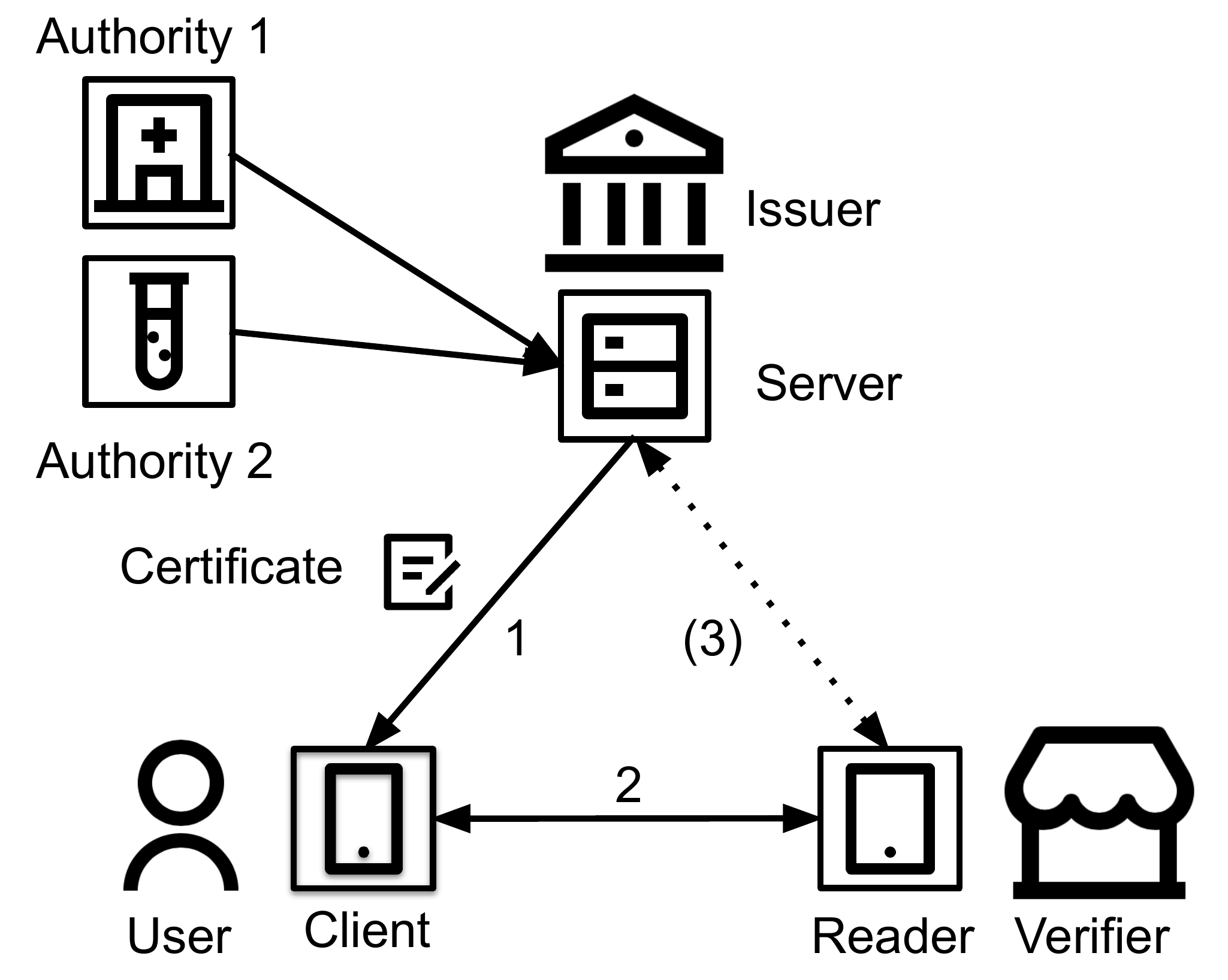}
	\caption{Illustration of the typical structure for DHP.}
	\ifnum\fullversion=0
	\Description{An illustration of the different parties taking part in a DHP flow. }
	\fi
	\label{fig:dhp}
\end{wrapfigure}
Despite the fact that solutions take different forms in different countries, to the best of our knowledge, the vast majority of DHPs follow the same overall idea, which we illustrate in Fig.~\ref{fig:dhp}. 

A user requests a \emph{certification} on their health status from an \emph{issuer} using their personal \emph{client} device. The client device can then present the certified health status towards a \emph{verifier} of the user's choice.
Concretely, the client will execute a protocol with the issuer's \emph{server} (optionally using information about the verifier gathered from the verifier's \emph{reader}), in order to retrieve a certification on the user's health status. 
The data used for constructing the certification is based on information the issuer has gathered from one or more \emph{health authorities} (with the user's consent). 
At the user's will, the client then presents information towards the verifier to allow it to verify the certification learned from the issuer. 
This can be done either visually (based on what is displayed on the client) or digitally by the client communicating with the reader.
Alternatively, to prevent reusing single-use information, the reader might validate the information received from the client by communicating with the server.

It is almost needless to say that a pure visual verification does not present much security against fake certifications given that a user is in control of the client and thus able to make it display any specific data. That is, they could make it simply display a message saying that they are validated to be safe, when this is not the case. 
Thus, for the rest of this paper, we assume some sort of interaction between the client and the reader is required to allow the verifier to accept the certification. 

To avoid the issue of revocation of a certification, we assume that any validation is time-constrained. That is, if a user validates towards a verifier, the issuer certifies that at this instance the user fulfills whatever health requirements are in place. 
This also means that we assume that the user is \emph{online} during validation, since they necessarily need to contact the issuer.
In section~\ref{sec:credentials}, we will discuss an alternative approach which works \emph{offline}, i.e. where the user does not need to talk to the issuer every time they validates towards a verifier. But for the other parts of this paper, we assume that a certification is time-constrained.

\paragraph{Security}
With this above requirement in mind, and inspired by the requirements of Hicks \emph{et al.}~\cite{HBMC20}, we can more concretely specify the security goals we want to realize:
\begin{description}
	\item[Correctness]~ If all parties are honest, the reader should accept the health status of the user controlling the client, in accordance with the data on the user being held by the health authorities. I.e. the information the verifier learns is authorized by the authorities and linked to the user controlling the client.
	\item[Forge-proof]~ A malicious user cannot\footnote{By \emph{cannot} we mean ``except with negligible probability'' and assuming that the underlying cryptographic assumptions used in the protocol, hold.} make the reader accept a health status which is not in accordance\footnote{By \emph{in accordance} we mean that the presented health status which a verifier learns about the user is \emph{derived} from the health information of that user being held by the authorities. This ensures that the user can be selective in how much information is shared, and hence enhance privacy.} with a certificate the issuer has sent to that user. That is, the user cannot manipulate the data contained in a certificate.
	\item[Binding]~ A malicious user cannot make the reader accept a health status that was issued to another user. I.e. the certificate must be linked to the specific user to whom it is issued. 
	\item[Unlinkability]~ One or more malicious readers cannot recognize whether the same user or distinct users have been validated in two different sessions.
	That is, the interaction between the client and the reader does not leave any ''link'' to the underlying user. This is the case both if the same user validates towards the same reader or the same user validates towards different readers at different verifiers. Thus, the verifiers cannot track when and where a specific user gets validated.  
	\item[Untraceability]~ A malicious issuer colluding with one or more malicious verifiers cannot trace a specific user to a specific validation. That is, the interaction between the client and the servers does not leave a ''trace'' that can be linked to the interactions between the client and the reader in the same protocol execution instance. Thus, it is not possible for the issuer to track which verifiers a specific user validates towards. 
\end{description}
Above, we discussed validation of a user through some client (practically being the user's app on their smart phone).
However, the issue still remains that the actual user should be linked to the digital identity for which the protocol is being carried out by the client. 
Otherwise it will be impossible to achieve \emph{protocol-level} binding.
This is unfortunately directly at odds with unlinkability/untraceability.
Since we focus on user privacy in this work, we will make a compromise towards the binding on the protocol-level, in order to achieve protocol-level unlinkability and untraceability. 
However, we discuss different hardening techniques that can be used to ensure \emph{deployment-level} binding between the user and the client in section~\ref{sec:hardening}.
This allows us to achieve a reasonable level of binding along with unlinkability and untraceability.



While both unlinkability and untraceability might seem minor, we will argue that they are not. 
Whether different verifiers collude to allow linkability, or a verifier and the issuer collude to allow traceability, both situations can happen without the user ever finding out! 
That is, there is no visible trace to the user that any of these situations has happened. 
Thus, the user will simply have to trust that this does not happen.
Consequently, the user will have to trust \emph{all} the verifiers \emph{and} the issuer in a manner where there will not even be a public digital trace if this trust is breached. 
Basically, the verifiers or issuer needs to be caught red-handed data-sharing in order to prove foul play has happened.
If the issuer is a government, this presents an issue – in particular in democracy-challenged countries where the government might exert extensive power over private businesses. 

Furthermore, if we consider a realistic deployment setting, we note that the issuer will also hold data from multiple health authorities on \emph{each} user. 
The reason is that it would otherwise be highly inconvenient and slow for a user to manually share certificated information from multiple health authorities each time the user wishes to get a certification. 
Hence, we imagine that the user consents to the issuer pulling this information automatically when needed, or getting updates automatically pushed from the authorities. 
Thus the issuer will not only have the ability to certify users' health attributes towards verifiers on-demand, but also hold a treasure trove of private information, even if unlinkability and untraceability are built into the protocol.  



\subsection{Distributing the Issuer}
\label{sec:distribution}
At Euro S\&P Baum \emph{et al.}~\cite{BFHLY20} introduced PESTO; ProactivEly Secure disTributed single sign-On. 
Their work shows how to increase user security in password authenticated Single Sign-On (SSO) solutions.
Their idea is to protect the user against compromised servers by introducing a cryptographic protocol for distributed password authentication.
This protocol works in such a way that all servers must be corrupt at the same time to even just allow for a \emph{brute-force} attack on the users' password. 
However, their approach also protects against corrupted servers by requiring \emph{all} the servers working together to issue a \emph{token}.
This token is basically an arbitrary message which is digitally signed using a \emph{distributed digital signature scheme}. 
That is to say a signature scheme where \emph{each} server holds a secret key share that must be used in order to construct a complete signature. 


We see that the idea of Pesto can translate to the setting of DHP, allowing us to realize a protocol with password-based authentication, where the issuer aggregates data from authorities and then certifies these in the shape of a time-constrained token. 
Thus simply using Pesto will greatly improve the overall security of DHP solutions where the issuer is centralized.

However, distribution does not purely solve problems but also create new ones: The user's sensitive data (vaccination dates/types, test results, and other data received from the health authorities) will now be located on multiple servers instead of a single one, which allows multiple points for attacks.
Furthermore, distribution does not directly help with issues of unlinkability or untraceability. 

\subsection{Our Contributions}
In this paper we show how to realize an unlinkable DHP solution based on Pesto and how simple \emph{secret sharing} can be used to ensure the confidentiality of the user's data at rest – even if some of the servers are malicious and colluding. 
Furthermore, we will show how replacing the distributed signatures used in Pesto with a \emph{blind} version can be used to achieve untraceability.
Finally, we discuss how commonly deployed hardening techniques, such as secure hardware, can be used to make it hard for malicious users to break the binding requirement.

\paragraph{Disclaimer}
This paper does not imply that the authors approve of the usage of DHPs as a tool, neither in the current, nor in future pandemics. This paper is only meant to suggest how to \emph{improve} the privacy and security of citizens whose governments have \emph{already} decided to use DHPs, \emph{without} taking the need for strong security and privacy into account.

\section{Related Work}
Despite the fact that corona-based DHPs have been put into commercial service only recently, a non-negligible body of research has already been carried out in this area.
Hicks \emph{et al.}~\cite{HBMC20} describe a general framework and features desirable in a DHP and how to realize these in either paper or app form. 
Their main idea is simply for the user to just present a signed certificate containing test or vaccination information, along with the user's picture and name, to a trusted verifier. 
While simple and efficient, it does not provide any unlinkability or untraceability. 
Another work by Butler \emph{et al.}~\cite{BHBMC20} suggests to use differential privacy when issuing a DHP. 
The DHP would contain an integer representing the health risk of the user, which is a differential private version of the user's true risk. That is, either their true risk or a randomly sampled risk.
On the macro-level their approach ensures that the overall risk of a \emph{cohort} of users represents the true risk.
While this prevents immunization discrimination in the macro setting, it is clear that this approach will not work in all contexts. In particular in settings where it is crucial to know the risk of a single, particular user. 
Furthermore, the issue of tracing and linking a user's movement is still an issue with this approach. 
It also applies that if the user's identity is not part of the token, there is no binding, for which reason a market for token sharing could arise.
If the tokens are issued on demand, the randomness of the issuance allows users to simple try to sample new tokens until they get a low-risk one.
Other solutions suggest to store test and vaccination data on the blockchain since the data will be immutable, in a way that can be checked by (trusted) verifiers~\cite{immupass,HSJA+20}.
Neither of these solutions provide unlinkability or untraceability. 

Finally, ISO has worked on a standard~\cite{eHealth} for DHPs based on the standard for Mobile Driver's License (mDL)~\cite{ISO18013-5}. Since an mDL is basically also a certification on certain attributes, it makes sense that the system for mDL can also be used for medical passports. 
The mDL standard offers both online and offline versions. One of the online versions is very similar to the general flow of the solution of Hicks \emph{et al.}~\cite{HBMC20}. 
Although the standard provides unlinkability, untraceability is not necessarily fulfilled. At least not in the online case. For the offline case, the standard can be realized using the alternative approach mentioned in Section~\ref{sec:credentials}.

\ifnum\fullversion=1
We note that this paper only presents a \emph{sketch} of how existing primitives can be used to improve privacy and security over existing medical passport solutions. We do no claim that these solutions are complete, nor that they are formally secure. We only argue on an intuitive level how such a solution could be constructed. The specific choices of components must be closely evaluated and proved to compose in a secure manner before these ideas are being deployed. We leave this as future work for anyone interested. 
\fi

\section{Preliminaries}
We use $x \getsr Y$ to denote that $x$ has been uniformly random-sampled from a set or procedure $Y$, and $x:=y$ to denote that $x$ gets assigned the value $y$. 
We let $[y]$ represent the size $y$ set of positive integers less than or equal to $y$. I.e. $[y]=\{1, 2, ..., y\}$.
We will use brackets, $[\cdot]_\cdot$ to denote hidden or randomized versions of variables. I.e. $[x]_r$ means some value constructed from $x$ using some randomness $r$ which means that for $b\gets \{0, 1\}$ given $[x_b]_r$ and $x_1$ and $x_2$ it is not possible for an algorithm running in polynomial time in a security parameter to recover $b$ with non-negligible probability if it has no further information on $r$.
We assume $\Hash(\cdot)$ denotes a cryptographic hash function, which is modelled as a random oracle. 

We assume the setting where the \emph{issuer} is realized by $\servers$ servers, with the $i$'th server, under an arbitrary ordering, denoted by $\server_i$. By server we here mean the program running on a physical machine. When needed, we will distinguish this from the \emph{issuer}, which is the legal entity controlling the servers. 
For each server we assume $\server_i$ has some private, high entropy, key information $k_i$. We note that internally $k_i$ contains several separate pieces of key information but we simply use $k_i$ for ease of abstraction.
We denote a \emph{client} by $\client$. When discussing multiple clients we use subscript, e.g. $\client_1$ and $\client_2$ to distinguish them. By a client we here mean the app being run on a specific \emph{user's} smart phone. We assume the user has a unique user ID, $\uid$, and a password $\pwd$.
Similarly, we denote a reader by $\reader$. When discussing multiple readers, we use subscript, e.g. $\reader_1$ and $\reader_2$ to distinguish them. By reader we here mean the app running on a specific device under the control of a specific \emph{verifier}. We assume that all readers have access to the same public key information $\pkissuer$ related to the key information held by the servers. 
We use the term \emph{party} to denote any of the $\server$, $\client$ or $\reader$ roles.
When we talk about an adversary, we mean an entity controlling a subset of the parties. That is, an adversary fully sees the internal state of these parties and controls how this subset of parties interact with the honest parties. We assume that the adversary is bounded by probabilistic polynomial time in a security parameter (which varies depending on the underlying algorithms).

\subsection{Secret Sharing}
\label{sec:secret-sharing}
Secret sharing is a secure approach to distributing some sensitive data $x$ to $\servers$ servers in such a way that less than $t\leq\servers$ colluding servers learn \emph{nothing} about the data $x$, but $t$ or more servers can completely reconstruct $x$.
A scheme achieving this is called a $(t, \servers)$ secret sharing scheme.
The idea was independently introduced by Shamir~\cite{Shamir79} and Blakley~\cite{Blakley79} in 1979, who showed how to do this efficiently for any $1<t\leq \servers$.
However, for our purpose we only need a $(\servers, \servers)$ secret sharing scheme which can be realized extremely efficiently by simply using XOR from the following folklore construction:
\begin{construction}
For secret $x\in\zo^\ell$ compute a $(\servers, \servers)$ secret sharing by selecting $x_1, \dots, x_{\servers-1} \getsr \zo^\ell$ and set $x_\servers=x \bigoplus_{i\in[\servers]} x_i$.
\end{construction}

\subsection{Digital Signatures}
\label{sec:signatures}
A standard signature scheme, $\SIG$, consists of the methods:
\begin{description}
	\item [$\SIG.\kgen( \normalfont{\aux} )\tor (\sk,\pk)$:]~ Randomized key generation algorithm which generates public key, $\pk$ and private key, $\sk$. It is required that $\aux$ contain randomness of at least the length of the security parameter.
	\item [$\SIG.\sign(\sk, m)\tor \sigma$:]~ Randomized signing algorithm which generates a signature, $\sigma$ on a message, $m$.
	\item [$\SIG.\vf(\pk, m, \sigma)\to \{\top, \bot\}$:]~ Deterministic verification algorithm which returns $\top$ if and only if $\top \gets \SIG.\vf(\pk, m, \allowbreak \SIG.\sign(\sk, m))$ for $(\sk, \pk)\gets \SIG.\kgen(\aux)$.
\end{description}
We require our digital signature scheme to be secure under \emph{Existentially Unforgeable under adaptive Chosen-Message Attacks (EU-CMA)}~\cite{STOC:GolMicYao83}. 
\ifnum\fullversion=1
This means that the signature scheme is secure if the adversary is given a public key $\pk$ generated with $(\pk, \sk) \gets \SIG.\kgen(\aux)$, yet cannot produce an arbitrary message $m'$ with a signature $\sigma$ s.t. $\SIG.\vf(pk, m', \sigma')$, \emph{even} if the adversary is also given access to a signing oracle that produces $\sigma\gets \SIG.\sign(\sk, m)$ for \emph{any} message $m$ of the adversary's choosing. The only obvious restriction is of course that $m'\neq m$.
\fi

\subsubsection{Distributed Signatures}
We furthermore require a \emph{distributed} version of digital signatures where the private key is secret shared between multiple parties so they must collaborate to construct a valid signature. This means that key generation now takes another parameter, $\servers$, reflecting the amount of shares of the private key to construct.
To ensure non-interactively between the parties running the algorithms, this means that the signature algorithm gets split in two: A \emph{partial signing} algorithm and a \emph{combining algorithm}.
Thus we define the \emph{distributed signature} $\DSIG$ as follows:
\begin{description}
	\item [$\DSIG.\kgen(\servers, \normalfont{\aux})\tor \left(\{\sk_i\}_{i\in[\servers]},\pk\right)$:]~ Randomized key generation algorithm which generates public key, $\pk$ and private key shares, $\{\sk_i\}_{i\in[\servers]}$. It is required that $\aux$ contain randomness of at least the length of the security parameter.
	\item [$\DSIG.\sign(\sk_i, m)\tor \sigma_i$:]~ Randomized partial signing algorithm that generates a partial signature, $\sigma_i$ on a message, $m$.
	\item [$\DSIG.\comb(\sigma_1, \dots, \sigma_\servers) \tor \sigma$:]~ Deterministic combining algorithm that aggregates the partial signatures, $\sigma_i$, and returns a single signature, $\sigma$.
	\item [$\DSIG.\vf(\pk, m, \sigma)\to \{\top, \bot\}$:]~ \sloppy Deterministic verification algorithm that returns $\top$ if and only if $\top \gets \DSIG.\vf(\pk, \allowbreak m, \DSIG.\comb\left(\DSIG.\sign(\sk_1, m), \dots, \DSIG.\sign(\sk_\servers, m)\right))$ for ${(\{\sk_i\}_{i\in[\servers]}, \pk) \gets \DSIG.\kgen(\aux)}$.
\end{description}
\ifnum\fullversion=1
Intuitively, we say this scheme is secure if it is EU-CMA and has the following two features as well:
\begin{description}
	\item[Share simulatability]~ An adversary learning at most $\servers-1$ key shares cannot distinguish whether the last share of a signature, $\sigma_i$, has been simulated using a random key share $\sk_i'$ (and knowledge of the real combined $\sigma$), or constructed using the real $\sk_i$. 
	\item[Signature indistinguishability]~ An adversary cannot distinguish signatures generated in a distributed way, compared to the ones computed by a single, true signing key $\sk$, aggregated from $\{\sk_i\}_{i\in[\servers]}$.
\end{description}
\else
Intuitively, we say this scheme is secure if it is EU-CMA secure, signature shares can be simulated, and the verification algorithm is agnostic to whether the signature was constructed in a distributed or non-distributed manner.
\fi

\ifnum\fullversion=1
\paragraph{Proactive Security}
If the private key shares $\sk_1, \dots, \sk_\servers$ can be randomized dynamically without affecting the public key $\pk$, we say that the scheme is \emph{proactively secure}.
For such a signature scheme the key generation algorithm also returns $\servers$ backup key shares $\{\bk_i\}_{i\in[\servers]}$ along with the regular ``live'' key shares $\{\bk_i\}_{i\in[\servers]}$. The backup key shares can be stored cold and offline, as these are only needed when the key shares are being refreshed. Specifically, the key shares are refreshed using a refreshing method\footnote{For technical reasons the signing algorithm is also required to take a nonce as input as well.}:
\begin{description}
	\item [$\DSIG.\refresh(\bk_i)\tor \sk_i'$]~ Randomized refresh method that construct a new key share $\sk_i'$ consistent with $\pk$.
\end{description}
A scheme is intuitively proactively secure if the adversary gets to decide continuously which key shares to learn but cannot break the underlying security of the signature scheme as long as they have at most $\servers-1$ key shares in between executions of $\DSIG.\refresh$.
\fi

\subsubsection{Blind Signatures}
Blind signatures were introduced by Chaum in 1983~\cite{C:Chaum82}.
A blind signature is basically a digital signature scheme where the signer is oblivious to the message being signed. 
This is achieved by having the party holding the message randomizing it in a special way before handing it to the signer. The signer then signs the randomized message and returns it to the message holder.
Finally, the message holder can then maul the signature, based on the randomness they used to randomize the message, in such a way that a valid signature is produced on the real and true message. 
A distributed version of blind signatures is defined in the natural way and were originally introduced by Desmedt~\cite{C:Desmedt87}. 
Regardless of whether the signature scheme is centralized or distributed, all existing algorithms remain the same for the blinded version.
However, we need introduce two new algorithms to allow it to be \emph{blind}, respectively \emph{unblind}:
\begin{description}
	\item [$\SIG/\DSIG.\blind(m)\tor (\beta, \tau)$:]~ Randomized blinding method that 
	constructs a blinded message $\beta$ from a plain message $m$, which also yields a trapdoor $\tau$ that can be used to derandomize a signature.
	\item [$\SIG/\DSIG.\unblind(\sigma, \tau)\to \sigma'$:]~ Deterministic derandomizing method that mauls $\sigma$ (a signature on $\beta$) to a valid signature $\sigma'$ on $m$ using trapdoor information $\tau$.
\end{description}
\ifnum\fullversion=1
Different definitions of security for blind signatures exist, but one of the most intuitive ones is the \emph{honest user unforgeable} definition~\cite{JC:SchUnr17}, which improves upon an earlier definition~\cite{C:JueLubOst97}.
Roughly this definition says that a blind signature scheme is secure if an adversary with oracle access to the signing algorithm, cannot produce a valid signature and message pair, which was not already queried to the oracle. 
This must also hold in the setting of a black box user being controlled by the adversary.
\fi

\subsection{Pesto}
\label{sec:pesto}
\begin{figure}[]
	\centering
	\begin{fboxenv}
	\begin{minipage}{0.5\textwidth}

\begin{center}
	\textbf{Enrolment Protocol} 
\end{center}
\begin{enumerate}
	\item The user inputs $\uid$ and $\pwd$ into $\client$.
	\item $\client$ picks some randomness $r$ and computes a hiding of $\pwd$, i.e. $[\pwd]_r$.
	\item \label{enum:aux-define} $\client$ then sends $(\uid, [\pwd]_r, \aux)$ to each $\server$, where $\aux$ is some auxiliary session information.
	\item Based on $(\uid, [\pwd]_r, k_i)$ each server $\server_i$ returns a value $y_i$ to $\client$.
	\item $\client$ aggregates $\{y_1, \dots, y_\servers\}$ using $r$ to retrieve a value $y$ which is \emph{pseudorandom} and uniquely and deterministically computed from $(\uid, \pwd, \{k_1, \dots, k_\servers\})$.
	\item Based on $y$ $\client$ computes $(\pkclient, \skclient) \getsr \SIG.\kgen(y)$.
	\item \label{enum:client-sign} $\client$ computes $\sigma_\client \getsr \SIG.\sign(\skclient, (\uid, \pkclient, \aux))$ and returns $(\sigma_\client, \pkclient, \aux)$.
	\item \label{enum:server-validation} Each server $\server_i$ ensures that $\top = \SIG.\vf(\pkclient, (\uid, \pkclient, \aux), \sigma_\client)$.
	\item Each server then defines $\aux_\client := (\uid, \pkclient, \aux)$ and compute $\sigma_{i, \aux} \getsr \DSIG.\sign(k_i, \aux_\client)$ and sends $\sigma_{i, \reg}$ to all the other servers.
	\item Each server $\server_i$ computes $\sigma_{\reg}\gets \DSIG.\comb(\sigma_{i, \reg}, \dots, \sigma_{\servers, \reg})$ which is the signature on $\aux_\client$.
	\item \label{enum:server-storage} Each server $\server_i$ verifies that $\sigma_{\reg}$ is correct by verifying $\top = \DSIG.\vf(\pkissuer, \aux_\client, \sigma_\reg)$. If so its it updates its associative data structure $U_i[\uid]:= \aux_\client$ and sends a registration acknowledgement to $\client$.
\end{enumerate}

\end{minipage}
\begin{minipage}{0.5\textwidth}

\begin{center}
	\textbf{Issuance Protocol} 
\end{center}


\begin{enumerate}
	\item The user inputs $\uid$, $\pwd$ and $m$ into $\client$.
	\item $\client$ picks some randomness $r$ and computes a hiding of $\pwd$, i.e. $[\pwd]_r$.
	\item \label{enum:client-iss-send} $\client$ then sends $(\uid, [\pwd]_r, m, \aux_{iss})$ to each $\server$, where $\aux_{iss}$ is some auxiliary session information.
	\item Server $\server_i$ retrieves $U_i[\uid]\rightarrow \aux_\client$ and from this $\pkclient$.
	Based on $(\uid, [\pwd]_r, k_i)$ each server $\server_i$ returns a value $y_i$ to $\client$.
	\item $\client$ aggregates $\{y_1, \dots, y_\servers\}$ using $r$ to retrieve a value $y$ which is \emph{pseudorandom} and uniquely and deterministically determined from $(\uid, \pwd, \{k_1, \dots, k_\servers\})$.
	\item Based on $y$ $\client$ computes $(\pkclient, \skclient) \getsr \SIG.\kgen(y)$.
	\item \label{enum:client-sign-issuance} $\client$ computes $\sigma_\client \getsr \SIG.\sign(\skclient, (\uid, \aux_{iss}, m))$ and returns $\sigma_\client$ to the servers.
	\item \label{enum:server-verify} Each server $\server_i$ ensures that $\top = \SIG.\vf(\pkclient, (\uid,\aux_{iss}, m), \sigma_\client)$. If so it compute $\sigma_i\getsr \DSIG.\sign(k_i, m)$ and sends $\sigma_i$ to $\client$.
	\item \label{enum:client-combine} The client computes $\sigma\gets \DSIG.\comb(\sigma_1, \dots, \sigma_\servers)$ which is the signature on the token $m$ and verifies that it is correct by verifying $\top \gets \DSIG.\vf(\pkissuer,  m, \sigma)$. If so its sends $(\sigma,  m)$ to $\reader$.
	\item \label{enum:verifier-check} $\reader$ verifies the token by validating $m$ against some external policy and verifying that $\top = \DSIG.\vf(\pkissuer,  m, \sigma)$.
\end{enumerate}

\end{minipage}
\end{fboxenv}
	\caption{The Pesto Protocol. A client $\client$ wishes to first enrol and later get arbitrary tokens issued.  The client interacts with with servers $\server =\{\server_1, \dots, \server_\servers\}$ where $\server_i$ holds key information $k_i$ and an initially empty associative data structure $U_i$. The client wishes to get a token verified by an arbitrary reader denoted $\reader$.}
	\ifnum\fullversion=0
	\Description{An somewhat concrete description of both the enrolment and issuance part of the Pesto protocol.}
	\fi
	\label{fig:pesto}
\end{figure}
We will not go into the underlying technicalities and formal security model of Pesto, as this is not needed to achieve the goal of this paper. Instead we refer the reader to the original Pesto paper for details~\cite{BFHLY20}. 
The overall idea of the protocol is that the client derives a pseudorandom value based on the user's id, $\uid$ and their (possibly low-entropy) password, $\pwd$ and the (high entropy) key share each server holds, $k_1, \dots, k_\servers$.
This is done using a protocol where each server does not learn anything only based on the entropy of the user's password. Instead everything they learn will have some high entropy sampled by the client mixed in as well. 
Hence it will not be possible to brute-force the user's password \emph{without} all the servers key shares. 
The client can then use the pseudorandom value to derive keys for a digital signature scheme. 
For enrolment the client shares the public key of this scheme with the servers and they store this in lieu of the user's (hashed) password.
For verification the client uses the signing key to sign a nonce, and shares this signature with the servers to prove that they used the same password as during enrolment.
The servers can then construct a distributed signature on any message the client supplied. 
The signed message then constitutes a token that the client can pass on to a reader.
We describe the Pesto protocol concretely in Fig.~\ref{fig:pesto} but with the underlying cryptography abstracted away.

\ifnum\fullversion=1
Pesto achieves the following security guarantees:
\begin{description}
	\item [Malicious security]~ Pesto remains secure even if the adversary controls the corrupted parties and has them send data that does not follow the protocol specification.
	\item [Adaptive security]~ Pesto remains secure even if the adversary gets to pick which parties to corrupt after the beginning of the protocol.
	\item [Proactive security]~ Pesto remains secure even if the adversary gets to decide continuously which parties to corrupt. I.e. the adversary can change which parties is has corrupted at a given instance. 
	\item [Dishonest majority]~ Pesto remains secure as long as the adversary only corrupts at most $\servers-1$ servers at a given time and \emph{any} amount of clients or verifiers.
	\item [Universally Composable (UC) security]~ The security of Pesto is proven in the UC model~\cite{FOCS:Canetti01} which is a very strong model of security, promising that it remains secure when used in composition with other protocols, as long as the underlying (distributed) signature scheme obeys certain security requirements. 
\end{description}
\fi
When we say that Pesto is secure, we intuitively mean that a client can only get a valid token from the servers if that client knows the password used during enrolment. Furthermore, the servers do not learn any non-negligible information about the honest user's password, under a Diffie-Hellman-like assumption. 
Similarly a reader \emph{only} accepts a token if \emph{all} servers agree that a token should be issued.
\ifnum\fullversion=0
This holds even if all servers, except one, are acting maliciously and colluding. 
\fi

\subsection{Secure Hardware}
\label{sec:secure-hardware}
Secure hardware is a valuable asset when it comes to real world practical security. 
Secure hardware allows hardening, and sometimes even full protection, against a malicious individual physically controlling the device where the secure hardware is located. 
The hardware could either be a Secure Enclave or Trusted Execution Environment (SE/TEE) on a phone, or a Hardware Security Module (HSM) in a server room. 
However, several attacks on secure hardware exist if the adversary has physical access to the hardware~\cite{NPSQ03}, or sometimes even access to running system level processes on it~\cite{USENIX:VMWGKPSWYS18}.
Still, secure hardware generally is apt as a practical hardening technique to prevent extraction of key material from a device \emph{even} if an adversary has gained \emph{limited} remote access to this device. 

Generally, we assume that secure hardware is able to store keys and perform standard cryptographic operations using these, through authorized system calls. 

For this paper, we assume that an adversary who is not the legitimate owner of the hardware cannot extract keys from the secure hardware, nor use it to perform cryptographic operations. 
However, we assume the owner can use it as they please, and we acknowledge the chance that with sufficiently large effort, they will be able to extract keys.

\section{Pesto Architecture for DHPs}
\label{sec:pesto-dhp}
Based on the description of Pesto in Fig.~\ref{fig:pesto}, we discuss how to use it to realize a DHP system based on the overall DHP architecture discussed in Section~\ref{sec:intro}.
During enrolment the user will provide the servers with consent to contact the health authorities holding the user's medical data.
Assuming this consent can be expressed as some data $c$, the user's client simply lets $c$ be part of $\aux$ which it sends in step \ref{enum:client-sign} of Enrolment.
The servers in turn validate $c$ as part of step \ref{enum:server-validation} and store this as part of $\aux_\client$ in their associative data structure in step \ref{enum:server-storage} in Enrolment.
At regular intervals the servers can then use $c$ to pull information from the health authorities.
Say the data from the health authority $i$ is denoted as $h_i$, the servers can then retrieve $U[\uid]\to \aux_\client$ and update $\aux_\client$ to also contain $h_1, \dots$ and store this back in the data structure $U$.

At some point Issuance will be executed. This happens when a user with ID $\uid$ goes to a verifier with ID $\vid$ in order to prove that their medical status fulfils the requirements put forth by the verifier. 
This is done by the user's client communicating with the verifier's reader. 
The reader returns a nonce, denoted by $q$, used to define a specific authentication session, and a timestamp of the current time, denoted by $t$.
It also returns an identifier of the verifier and the specific reader, denoted by $\vid$.
Finally, it also returns the specific requirements the verifier requires the user  to fulfil.
We call these requirements the \emph{public policy}. This could for example be that the user's vaccination must be at most 6 months old or that they must have tested negative within 24 hours. 
We denote the the public policy by $\pp$.
If the user accepts the level of detail of the public policy, i.e. they accept that the verifier learns that they fulfil this, then they enter their password into the client.
In step \ref{enum:client-iss-send} of Issuance, the client will then construct a message $m$ consisting of $(\pp, \vid, q, t)$.
In step \ref{enum:server-verify} the servers then verify that $\pp$ is indeed fulfilled in accordance with the data $h_1, \dots$ they hold in $\aux_\client$ and that $t$ contains the current time (or is reasonably close to it).
The user then gets a signature on $m$ and returns it to the reader. 
In step \ref{enum:server-verify} the reader then checks the signature on $m$ and that it contains $(\pp, \vid, q, t)$ that it originally sent to the client and that $t$ is still reasonably close to the current time. 

It is clear to see that the above protocol yields a distributed version of DHP, with the advantage that servers must all agree on making a token (i.e. $m$ and an associated signature).
This approach also prevents linkability since no user-specific information is contained in the token. However, this means that it is possible for the user to share this token with another user using authorizing towards the same verifier at the same time. 
It is reasonable to ensure that a token can only be used once by the reader (by simply only allowing the execution of one session at a time). 
However, this still does not prevent a \emph{relay} attack breaking the binding of the system. 
In such an attack one malicious user, who fulfils $\pp$, allows another user to use their authentication towards the issuer to get a token. 
We will discuss this problem further in Section~\ref{sec:deployment} since this is an inherent problem if we still want unlinkability. 

\subsection{Removing Traceability}
The protocol outlined above still allows for traceability through the signature on the token given to the verifier.
However, if the public policy is always the same
then we can remove traceability having the server make a blind signature on \emph{any} message the user requests. 
That is, instead of only using a distributed signature scheme, we use a blind distributed signature scheme.
Like before, when a user with ID $\uid$ wants to prove that they fulfil the requirements of some verifier, and their client gets $(\vid, q, t)$ from the reader.
Now instead of sending $m$, containing $(\vid, q, t)$, the client computes $\beta$ using $\DSIG.\blind(m) \tor (\beta, \tau)$ and sends $\beta$ instead of $m$. 
The servers then proceed as described above, but with $\beta$ instead of $m$.
In step \ref{enum:client-combine} of Issuance, the client then computes $\DSIG.\unblind(\sigma, \tau) \to \sigma'$ and use $\sigma'$ when for the rest of the protocol instead of $\sigma$.
The crucial observation being that the servers never used the content of $m$, neither in Pesto nor in our protocol.

If we wish to allow for multiple possible public policies, the servers need a different distributed signing key pair for each of the possible public policies we will allow.
That is, each distributed signing key pair is associated with a specific policy.
This means that the reader can verify if the user fulfils the correct policy by simply ensuring that the expected signing key, associated with this policy, was used to construct the signature.

\subsubsection{Security}
We first note that the token now does not contain any information traceable to the user (assuming enough users are getting tokens of the specific policy issued within the granularity of the timestamp). 
Furthermore, the randomness added by the blinding ensures that the servers will not be able to recognize the signature it made for the user with $\uid$ \emph{even} if they see the unblinded version. E.g. if the verifier shares the tokens with the issuer. 

Next, we see that despite being allowed to get \emph{anything} signed, this new approach does not give any advantage to a user to get a malicious token constructed. 
The reason is that a user must still authenticate towards the servers first, \emph{and} have the servers verify that they fulfils the public policy before signing a token. 
Still, as a malicious user could get anything signed, they could request a signature on a future-dated token. However, since the token contains a nonce, specified by the reader, this will not help a malicious user since the client only learn that nonce once they start verification towards the reader. 
Thus, the only token the user can get constructed \emph{and} accepted at \emph{any} reader is one linked to that specific reader, at that specific time, for the current session in progress at that reader. 
This ensures that the approach is forge-proof if the specific reader only allows for one verification session at the time.



\subsection{Preventing Sensitive Data Replication}
As previously discussed in Section \ref{sec:distribution}, the current approach increases the visibility of sensitive user data since the data from \emph{all} the issuing authorities will be stored locally on all the servers, which is in contrast to a centralized approach where the data is just stored in a \emph{single} centralized server.
We could imagine solving this by only fetching data on-demand from the medical authorities when a user requests a token.
However, this can provide congestion and unnecessary communication.
To prevent this, we instead have the servers store all the relevant up-to-date medical data locally. 
To ensure that an attack on \emph{any} of the servers does not leak private information, the servers will secret share the data with each other. 
Let $h\in\zo^\ell$ denote the authenticated health data on a user the servers receive from the health authorities. 
Once receiving this (be it through a push or a pull operation with the health authorities) each server $i\in[n-1]$ forms an xor secret sharing by picking a random value $x_i\gets \zo^\ell$ and sending this to server $n$. 
Server $n$ then computes $x_n=h \bigoplus_{i\in[n-1]} x_i$. 
Finally, each server $i$ also computes $d\gets \Hash(h)$ and stores $(d, x_i)$ as part of $\aux_\client$ stored in $U[\uid]$, where $\Hash (\cdot)$ is a cryptographic hash function. 

Once a user with $\uid$ requests a token from the servers, then server $i$ retrieves $(d,x_i)$ from $U[\uid]\to \aux_\client$ and sends $x_i$ to every other server.
Thus they can each restore $h'=\bigoplus_{i\in[n]}x_i$ and verify that $h'=h$ by ensuring that $\Hash(h')=d$, which prevents the possibility of a malicious server manipulating the data. 

The servers then use $h$ directly from the RAM to verify whether the policy of the user is obeyed at that given time. Once the request has been verified, the servers delete $h$ from the RAM.
This approach ensures that the servers don't store private user data in plain at rest, and that they can only restore a user's data once \emph{all} servers agree that this is okay to do. 

\section{Deployment}
\label{sec:deployment}
\subsection{Deployment Security}
\ifnum\fullversion=1
We have discussed how to use an augmented version of the Pesto protocol to facilitate DHPs with a focus on user-privacy through unlinkability and untraceability, but at the cost of not obtaining protocol-level binding of the user.
However, we have not discussed how to actually deploy this, and in turn, how to achieve deployment-level binding through different hardening techniques. 
\fi

First we note that the current system does not consider, nor prevent, an illegitimate user from enrolling. 
Still, this can be ensured using standard techniques and is the tangent problem of linking a physical identity to a digital identity. 
This has been considered by many institutions, so we will simply assume that enrolment is linked to a suitable solution~\cite{eidas,ISO24760}. 

Thus the main issue is to ensure that the person using a client is bound to the user for whom the issuer constructs a token.
This this means we must prevent two issues; \emph{token-sharing} (sharing a specific token) and \emph{client-sharing} (a user sharing a client linked to them with another user).
Considering token-sharing, it seems we have already prevented this through linking the token to the nonce and timestamp of the reader. Unfortunately this is not enough.
Since the token does not (and should not) contain any information linking it to the user for which it is issued, there is nothing preventing two users colluding to pretend to be a single user, by having the clients relay data between reader and issuer respectively.
More specifically, user 1's client initiates the protocol with a reader and thus receives $(\vid, q, t)$.
The client then transmit this to user 2's client. 
User 2's client then requests a token on a message containing $(\vid, q, t)$ from the issuer.
User 2's client then sends the signature $\sigma'$ to user 1's client. 
User 1's client can now present a valid token to the reader without having been authenticated and validated by the issuer!
Still, since the token can only be used \emph{once} and must be constructed at the instant at which it should be used, this attack does not give any advantage over a simpler client sharing attack, where user 2 simply gives their client to user 1 along with their password.
The low-tech approach to prevent such misuse, as for example seen in national IDs and passports, is to include a picture or physical description of the user to be validated. 
Of course this info cannot be included in the token itself as it would otherwise allow for linkability and traceability. 
Thus the only solution is to have such information be displayed on the user's client, hence making it function almost like a physical ID card. 
It is crucial to actually use a picture rather than and a unique ID linkable to the individual. 
The reason being that a skilled verifier can easily remember such an ID and note it down as soon as the user has been verified, thus breaking linkability and traceability. 
This is not feasible with a picture, as the only way to identify a user from memory of a picture is by looking through all potential pictures. 
Still, it is crucial that no camera is pointed at the user's device \emph{when} the picture is shown, as the user could otherwise be identified with reverse image search. 

\subsection{Hardening}
\label{sec:hardening}
Since the user owns their own device, we must assume they are able to manipulate it. 
Thus simply including a picture on the client as part of the verification is not sufficient to prevent client or token sharing.
We can for example imagine that an overlay app can present an incorrect picture, while the token communicated with the reader is the legitimate one. 
Several hardening techniques exist that can help deter these problems by strengthening the binding (of a user to their client).
For example the following standard app techniques can be used to prevent construction and usage of malicious apps which could allow fake pictures being showed (allowing device-sharing) or constructing apps that could act as relays (allowing token-sharing):
\begin{description}
	\item[Runtime Application Self­ Protection (RASP):]~ This can be used ensure that an app is not running on a rooted or jail-broken device. It can also be used to detect source code modification at runtime and detect whether the app is running a real device or in an emulator. All of this makes it hard to modify a legitimate app and prevent malicious hooking (which could be needed to visually modify the user's picture being displayed).
	\item [Software obfuscation\footnote{Not to be confused with cryptographic obfuscation~\cite{FOCS:GGHRSW13}.}:]~ Provides limited protection against reverse engineering of the app's source code. This makes it harder for an adversary learning the source code and thus implement an overlay or illegitimate app.
	\item [API hardening:]~ iOS offers the \texttt{DeviceCheck} feature which can be used to verify that API calls come from the app they are expected to come from. Android has a similar feature in SafetyNet. This ensures that a maliciously constructed app is not making API calls. In particular this is critical since a malicious app could be used to construct the attack of two colluding user's as mentioned above.
\end{description}
To highlight the importance of hardening, keep in mind that when it comes to physical ID cards and certificates, they can also be copied, modified and borrowed. But doing so requires work \emph{each} time it happens. 
Whereas a solution to, say, fake a picture in an app, can be distributed digitally and in an instant, without overhead, and then be available to millions of users. 
Thus a technique for digital compromise can be scaled much easier than a technique for compromising physical cards. 

\tore{todo something about using a distributed signature for code signing}

\subsubsection{SE/TEE hardening}
A bit of an orthogonal hardening issue is the fact that a user \emph{only} needs the password used at enrolment in order to authenticate and get a token on \emph{any} device with the legitimate client app installed.
We can of course argue that since the issuer knows exactly which user is trying to request a token, how often this user has requested tokens and whether token issuance is successful, they can implement rate limiting and account closing if anything suspicious is going on. 
Even so, since none of the servers \emph{ever} see what the user picks as password, they cannot ensure that the user picks a password in accordance with a policy of sufficient length and mix of characters, and which is not known to be compromised. 
Still, since the password is supplied by the user to the client, it can locally verify that it fulfils the password policy and securely check it for compromise using \url{haveibeenpwned.com}.
However, there is nothing to do about a user sharing their password or using it at other sites which might get compromised, which happens very frequently~\cite{ponemon}.
In particular the password only constitutes a single factor of authentication, meaning the solution as described so far is insufficient for digital authentication e.g. in the EU~\cite{enisa}, except at the lowest legal level. 

Fortunately, if the client is an app on a modern iOS or Android phone, it is easy to add an additional factor, by levering the SE, respectively TEE, on the phone.
Specifically the SE/TEE offers secure hardware for storing app-specific keys and performing basic cryptographic operations with these. 
In order to leverage such secure hardware, the client will use it to construct a key pair, denoted $(\skse, \pkse)$ for digital signatures during enrolment, where the secret key, $\skse$ will be stored and never leave the SE/TEE. 
The public key, $\pkse$, on the other hand, will be shared with the servers in step \ref{enum:aux-define} of Enrolment. 
In step \ref{enum:client-sign} the client also computes $\sigma_\SE \getsr \SIG.\sign(\skse, (\uid, \pkclient, \pkse, \aux))$ and sends this to the server along with $(\sigma_\client, \pkclient, \pkse, \aux)$.
In step \ref{enum:server-validation} the servers then also verify that $\top= \SIG.\vf(\pkse, (\uid, \pkclient, \pkse, \aux), \sigma_\SE)$.
This binds the $(\skse, \pkse)$ key pair of the SE/TEE to the user enrolling. 
Now when the user runs the issuance protocol, they will not just use $\skclient$ for signing but also $\skse$. 
This allows the servers to ensure that the \emph{same} client as used for enrolment is used during issuance. 
Specifically in step \ref{enum:client-sign-issuance} the client also computes $\sigma_\SE \gets \SIG.\sign(\skse, (\uid, \aux_{iss}, \beta))$ and sends this to the servers.
Then in step \ref{enum:server-verify} the servers also verify that $\top = \SIG.\vf(\pkse, (\uid, \aux_{iss},\beta), \sigma_\SE)$.

Because the SE/TEE is secure hardware it means that the key pair offers an extra factor in \emph{possession}, whereas the password needed during authentication offers a factor in the form of \emph{knowledge}. 
Of course a bad user can share their password, or spend a non-negligible effort in extracting the secret key from the SE/TEE, but this is not what we are trying to protect against here.

\ifnum\fullversion=1
\subsubsection{Server Hardening}
We remark that most real-world infrastructures implementing security critical server-based signing use secure hardware in the form of an HSM. 
This is to protect against theft and illegitimate use of their signing keys. 
Although it might not be too hard for an administrator to load malicious software onto the server, allowing \emph{temporary} misuse of the HSM, companies generally have procedures in place to catch such behaviour quickly. 
In our solution, no full keys are stored on an online server, but only key shares instead.
However, if these key shares are stored in plain, without the usage of the HSM, it is not too hard for a malicious administrator to steal and misuse these without any log or visible indication that this has happened. 
Thus it would be highly desirable to also have these be stored and used by an HSM.
Fortunately some HSMs, such as the Thales nSheild, allow the loading of custom firmware to allow for arbitrary computations on key material. 
Thus using such an HSM will allow servers to still keep the security guarantees offered by HSMs on top of the increased security achieved through distribution.
\fi

\subsection{External Factors}
\ifnum\fullversion=1
We have discussed how to ensure unlinkability and untraceability both at the protocol and deployment level.
However, to offer full privacy for the users, we still need to consider the infrastructure on which the system is used.
\fi

First of all, we must ensure that there is no identifiable link in the way the client communicates with the reader.
Since the interaction between the two devices only consists of one round of communication, this could be handled with QR codes on the client's and reader's screens.
This would ensure that no meta information is communicated.
However, such an approach can be a bit cumbersome and annoying from a usability standpoint. 
Thus Bluetooth might be used instead. 
Since Bluetooth devices have a unique MAC address, this poses an issue of linkability.
Fortunately with Bluetooth Low Energy (BLE), MAC address randomization has been introduced, ensuring that MAC addresses will be different in different interactions~\cite[Vol. 1, Part A, Sec. 5.4.5]{bluetooth} (at least if we make the reasonable assumption that a specific client is not executing the issuance protocol multiple times within a few minutes).

What poses a greater issue is the communication between the client and the server, since the IP of the client can leak a lot of information. 
In particular, if the client is using an Internet connection supplied by the verifier, e.g. through free WiFi, then the IP can be directly linked to where the user is executing the issuance protocol from.
This can be eliminated to a large extent by using standard anonymity techniques such as the Tor network or a trusted VPN, outside of the jurisdiction of the issuer. 

However, if the user is using a mobile data connection, their phone company will have information on which antenna their client is using during issuance. 
Thus, a collusion between the issuer and phone company could be used to link a specific user to an approximate location during issuance.
It is crucial to notice that even using the Tor network, or a trusted VPN, does not prevent this, as the mobile data provider will necessarily know the identity of the client whose data connection is being used. 
This issue could be prevented by using a prepaid SIM card which does not contain a linking to the specific user in question. 
Using such a SIM card, along with the Tor network, or a VPN, should provide pretty extensive anonymity of the user, assuming their client device regularly uses the SIM for an arbitrary data connection.

Even though it is possible to get decent anonymity even against an issuer colluding with the user's phone company, there is still a major anonymity issue left.
That is the anonymity against a colluding issuer and device manufacturer. 
It was recently shown~\cite{Leith21} that both iOS and Android share location data every few minutes with Apple, respectively Google. 
Thus if Apple or Google were to collude with the issuer, they can track approximately where the user is when they request issuance. 
Unfortunately this seems to be an inherent issue with any solution using an Apple or Google smart phone and requiring the user to interact with the issuer each time they need a token issued. 
The best approach to somewhat prevent this is to use a burner device and burner account with Apple or Google which does not involve any identifiable information of the real world identity of the user. 
However, the tracking will most likely still be linkable to a specific individual since Apple and Google will quickly learn the user's home address and work location through the device's location data. 
If the issuer has access to this information, it would be easy to even link the burner device and account to a specific individual.

\subsection{An Alternative Approach}
\label{sec:credentials}
Our \emph{online} solution which requires a client to contact the issuer when they wish to prove they fulfil the health requirements of the verifier is not the only possible approach to realizing DHPs. 
Specifically, by weakening the binding a bit further, it is possible to construct an \emph{offline} solution where the servers construct a \emph{credential} containing the data from the health authorities and sharing this with the client.
The client then stores this credential \emph{offline}, and when the user wants to convince the verifier that they fulfil a public policy, they construct a proof that they hold a credential, signed by the issuing authority, containing values that fulfil the policy. 
It is possible to ensure a limited lifetime of the credential by the issuer embedding timestamps into it. 
Constraints on the timestamps will then be part of the public policy such that the verifier can ensure that the health attributes are still valid, according to the public policy, at the time of verification. 
Thus a solution based on credentials can be used even when the user does not have an Internet connection, or issuer is not available. 
This also limits the amount of privacy leakage that can result from the device manufacturer or the user's phone company sharing location information on the client with the issuer.

Work on standardizing usage of offline credentials (although not necessarily generated in a distributed manner) is being carried out by W3C~\cite{w3c}.
Furthermore, the use of distributed credentials have also been suggested~\cite{SCN:CDLNT20} and it has also been discussed how to integrate these with Pesto in lieu of distributed signatures~\cite{MRLBS20,MBRFSMSPS20}.

While both credentials and their proofs can be efficiently realized, they have several liabilities compared to simple tokens using distributed signatures:
\begin{description} 
	\item [Client heavy computation:]~ Despite being practically efficient, more computation must still be carried out. In particular the computational requirements of the client become much more significant due to the need for constructing the relevant proofs to show to the verifier.
	\item [Illicit copying:]~ Since the credential can be used for multiple validations towards multiple verifiers, it means that nothing prevents the user from copying the credential and sharing it with other users. The other users can then falsely convince a verifier that \emph{they} hold a valid credential.
	Even though this is also possible with the tokens, as we discussed in Section~\ref{sec:deployment}, the fact that the tokens are only valid for a very short time, linked to a specific reader and single-use, it makes it much harder to share a token. 
\end{description}

\tore{Should I mention zk approach for hiding data?}
\tore{MPC could also be used but then I am cannibalizing other results.}
\tore{linking to external id based on random sample of identifier. ie. few digest of cpr}

\ifnum\fullversion=1
\begin{figure}[]
	\centering
	\begin{fboxenv}
	\begin{minipage}{0.5\textwidth}

\begin{center}
	\textbf{Enrolment Protocol} 
\end{center}
\begin{enumerate}
	\item The user inputs $\uid$ and $\pwd$ into $\client$.
	\item The client uses its SE/TEE to construct a key pair $\skse, \pkse$
	\item $\client$ picks some randomness $r$ and computes a hiding of $\pwd$, i.e. $[\pwd]_r$.
	\item $\client$ then sends $(\uid, [\pwd]_r, \pkse, \aux)$ to each $\server$, where $\aux$ is some auxiliary session information.
	\item Based on $(\uid, [\pwd]_r, k_i)$ each server $\server_i$ returns a value $y_i$ to $\client$.
	\item $\client$ aggregates $\{y_1, \dots, y_\servers\}$ using $r$ to retrieve a value $y$ which is \emph{pseudorandom} and uniquely and deterministically computed from $(\uid, \pwd, \{k_1, \dots, k_\servers\})$.
	\item Based on $y$ $\client$ computes $(\pkclient, \skclient) \getsr \SIG.\kgen(y)$.
	\item $\client$ computes $\sigma_\client \getsr \SIG.\sign(\skclient, (\uid, \pkclient, \pkse, \aux))$ and $\sigma_\SE \getsr \SIG.\sign(\skse, (\uid, \pkclient, \pkse, \aux))$ (using the SE/TEE) and returns $(\sigma_\client, \sigma_\SE, \pkclient, \pkse, \aux)$.
	\item Each server $\server_i$ ensures that $\top = \SIG.\vf(\pkclient, (\uid, \pkclient, \pkse, \aux), \sigma_\client)$ and $\top = \SIG.\vf(\pkse, (\uid, \pkclient, \pkse, \aux), \sigma_\SE)$.
	\item Each server then defines $\aux_\client := (\uid, \pkclient, \pkse, \aux)$ and compute $\sigma_{i, \aux} \getsr \DSIG.\sign(k_i, \aux_\client)$ and sends $\sigma_{i, \reg}$ to all the other servers.
	\item Each server $\server_i$ computes $\sigma_{\reg}\gets \DSIG.\comb(\sigma_{i, \reg}, \dots, \sigma_{\servers, \reg})$ which is the signature on $\aux_\client$.
	\item Each server $\server_i$ verifies that $\sigma_{\reg}$ is correct by verifying $\top = \DSIG.\vf(\pkissuer, \aux_\client, \sigma_\reg)$. 
	\item For $i\in[n]$, each server $i$ picks $x_i\getsr \zo^{|\aux_\client|}$ and sends this to server $n$. Server $n$ then defines $x_n:=\aux_\client \oplus \sum_{i\in[n-1]} x_i$.
	\item All servers compute $d:=\Hash(\aux_\client)$ and update their associative data structure $U_i[\uid]:= (x_i, d)$ and sends a registration acknowledgement to $\client$.
\end{enumerate}

\end{minipage}
\begin{minipage}{0.5\textwidth}

\begin{center}
	\textbf{Issuance Protocol} 
\end{center}


\begin{enumerate}
	\item The user starts $\client$ and the verifier starts $\reader$.
	\item $\reader$ samples $q \getsr \zo^{128}$ and lets $t$ be the current time and shares $(\vid, q, t)$ with $\client$.
	\item $\client$ receives $m=(\vid, q, t)$ from $\reader$ and ensures that $t$ is reasonably close to the current time and asks the user to authenticate.
	\item If the user accepts sending the request then they input $\uid$ and $\pwd$ into $\client$.
	\item $\client$ picks some randomness $r$ and computes a hiding of $\pwd$, i.e. $[\pwd]_r$, along with $\DSIG.\blind(\vid, q, t) \tor (\beta, \tau)$.
	\item $\client$ then sends $(\uid, [\pwd]_r, \beta, \aux_{iss})$ to each $\server$, where $\aux_{iss}$ is some auxiliary session information.
	\item Server $\server_i$ retrieves $U_i[\uid]\rightarrow (x_i, d)$ and sends $x_i$ to every other server.
	\item Every server then computes $\aux_\client=\sum_{i\in [n]} x_i$ and verifies that $d=\Hash(x_i)$, and from $\aux_\client$ retrieves $\pkclient$ and $\pkse$.
	\item Based on $(\uid, [\pwd]_r, k_i)$ each server $\server_i$ returns a value $y_i$ to $\client$.
	\item $\client$ aggregates $\{y_1, \dots, y_\servers\}$ using $r$ to retrieve a value $y$ which is \emph{pseudorandom} and uniquely and deterministically determined from $(\uid, \pwd, \{k_1, \dots, k_\servers\})$.
	\item Based on $y$ $\client$ computes $(\pkclient, \skclient) \getsr \SIG.\kgen(y)$.
	\item $\client$ computes $\sigma_\client \getsr \SIG.\sign(\skclient, (\uid,\aux_{iss}, \beta))$ and $\sigma_\SE \getsr \SIG.\sign(\skse, (\uid, \aux_{iss}, \beta))$ and returns $(\sigma_\client, \sigma_\SE)$ to the servers.
	\item Each server $\server_i$ ensures that $\top = \SIG.\vf(\pkclient, (\uid, \aux_{iss}, \beta), \sigma_\client)$ and $\top = \DSIG.\vf(\pkse, (\uid, \aux_{iss}, \beta), \sigma_\SE)$ and the client fulfils the public policy. If so it compute $\sigma_i\getsr \DSIG.\sign(k_i, \beta)$ and sends $\sigma_i$ to $\client$.
	\item The client computes $\sigma\gets \DSIG.\comb(\sigma_1, \dots, \sigma_\servers)$ followed by $\DSIG.\unblind(\sigma, \tau) \to \sigma'$.
	\item $\client$ then verifies that $\top= \DSIG.\vf(\pkissuer, m, \sigma')$. If so its sends $(\sigma', m)$ to $\reader$.
	\item $\reader$ verifies the token by validating $m=(\vid, q, t)$ according to what it initially sent, and that $t$ has not expired. Finally it also verifies that $\top = \DSIG.\vf(\pkissuer, m, \sigma')$.
\end{enumerate}

\end{minipage}
\end{fboxenv}
	\caption{The Full Protocol for DHPs. A client $\client$ wishes to first enrol and later get DHPs issued.  The client interacts with with servers $\server =\{\server_1, \dots, \server_\servers\}$ where $\server_i$ holds key information $k_i$ and an initially empty associative data structure $U_i$. The client wishes to get a token (DHP) verified by an arbitrary reader denoted $\reader$.}
	\ifnum\fullversion=0
	\Description{An somewhat concrete description of both the enrolment and issuance part of the full protocol.}
	\fi
	\label{fig:full}
\end{figure}
\fi
\section{Conclusion}
\begin{wrapfigure}{r}{0.45\textwidth}
	\centering
	\includegraphics[width=0.4\textwidth]{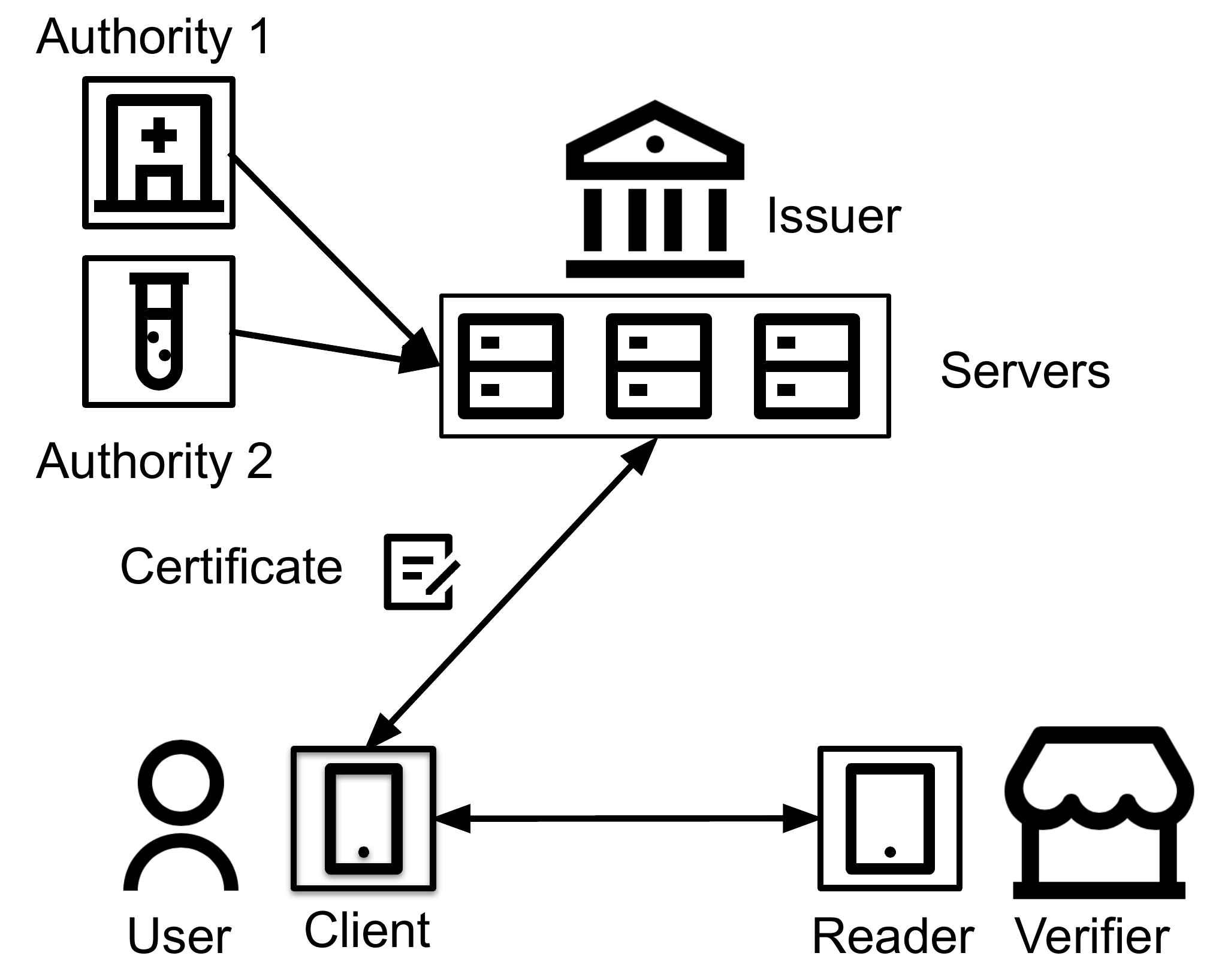}
	\caption{Illustration of our DHP structure.}
	\ifnum\fullversion=0
	\Description{An illustration of the different parties taking part in our DHP flow.}
	\fi
	\label{fig:ourDHP}
\end{wrapfigure}
We have showed how to make a simple and efficient protocol for DHPs based on multi-factor authentication that is both forge-proof, unlinkable, and untraceable on the protocol level and discussed how this can be deployed to also offer a reasonable level of binding. 
We also discussed how to increase security of the issuer and the privacy of the users' health data by using the distributed protocol Pesto and simple secret sharing. 
We illustrate the overall architecture of our protocol in Fig.~\ref{fig:ourDHP}.
\ifnum\fullversion=1
For completeness, we also give a semi-formal description of the entire protocol in Fig.~\ref{fig:full}, although abstracting away the issue of digital identity verification and the health authorities since all of these are highly deployment specific.
\fi
Finally, we have discussed which external factors could come in play to ensure full end-to-end unlinkability and untraceability for the user.

Our solution only uses a single round of communication between the client and the reader and only two rounds between the client and the servers. 
The data transmitted is minimal, in particular it is bounded by the security parameter plus the size of the representation of the message being verified by the reader.

Computationally we only require client, servers and verifier doing a constant amount of exponentiations (and other symmetric primitives).

\paragraph{Currently Deployed DHPs}
We note that legally speaking, some cross-country regulation and specification of DHPs are already in place. 
In particular the EU requires verification of a DHP to consist of validating a digitally signed long-term token encoded as QR code, containing both name and birthday of the user it has been issued towards, along with time-stamps and location information for testing and vaccination~\footnote{\url{https://ec.europa.eu/commission/presscorner/detail/en/qanda_21_1187}}.
This has the advantage of working fully offline, even without the user holding an electronic device. Of course this is also completely linkable and traceable. 

However, European countries also implement their own solutions along this.
In Denmark for example~\cite{coronapas}, this solution is used alongside time-constrained and signed\footnote{To the best of our knowledge a regular, non-blind signature scheme is used.} token that simply presents the verifier with a bit indicating whether the user is ``corona-safe'' or not. 
While this solution manages to obtain unlinkability, it still has the issue of traceability, due to the fact that the signature on a specific token, will be associated with a specific user, since the user has to authenticate before the issuer signs it.

Our solution could be used in the same manner as the Danish hybrid approach. That is, as a highly private online solution, with a less private offline solution, using the EU QR code. 
This means that the linkability and traceability is only possible in a minority of uses, and thus that the information that can be learned from this becomes significantly less valuable. 

\paragraph{Future work}
We remark that this paper has only considered the abstract realizations of the underlying cryptographic primitives. Thus future work includes specifying, and proving secure, a concrete choice of distributed blind signatures that fit into the proof of security of Pesto.

Further work includes investigating the use, and formal security, of \emph{partially blind distributed signature scheme} in the setting of Pesto. 
Concretely a partially blind signature scheme will allow the issuer to include the public policy in the message they sign.
This will allow usage of arbitrary policies with only a \emph{single} signing key associated with the issuer, or a limited amount of policies, each based on a specific issuer-public key.

\ifnum\submission=0
\ifnum\fullversion=0
\begin{acks}
\else
\paragraph{Acknowledgements.}
\fi
The author would like to thank Evangelos Sakkopoulos, Nuno Ponte and Michael Stausholm for useful discussions.
This work received funding from the EU Horizon 2020 research and innovation programme under grant agreement No 786725 OLYMPUS.
\ifnum\fullversion=0
\end{acks}
\fi
\fi

\ifnum\fullversion=0
\bibliographystyle{ACM-Reference-Format}
\else 
\bibliographystyle{alpha}
\fi
\bibliography{bib/abbrev2,bib/crypto,bib/additional}


\end{document}